# Enhancing User Resilience Against AI-Augmented Phishing: A Two-Stage Framework for Detection and Personalized Training

*Weihao Qu*
*Computer Science and Software Engineering Department*
*Monmouth University*
*West Long Branch, New Jersey, USA*
*email: wqu@monmouth.edu*
*ORCID:0000-0003-1027-6556*

*Gurmeet Singh*
*Computer Science and Software Engineering Department*
*Monmouth University*
*West Long Branch, New Jersey, USA*
*email: gsingh@monmouth.edu*
*ORCID:0000-0003-3800-5184*

*Daniel Crawford*
*Computer Science and Software Engineering Department*
*Monmouth University*
*West Long Branch, New Jersey, USA*
*email: s1323702@monmouth.edu*
*ORCID:0009-0004-0649-5355*

*Bingjun Li*
*Education Department*
*Monmouth University*
*West Long Branch, New Jersey, USA*
*email: s1363772@monmouth.edu*
*ORCID:0009-0006-5984-0364*

*Jalen Smith*
*Computer Science and Software Engineering Department*
*Monmouth University*
*West Long Branch, New Jersey, USA*
*email: s1358642@monmouth.edu*
*ORCID:0009-0007-9805-1744*

***Abstract***—The rapid development of artificial intelligence, including agents and deepfake techniques, has accelerated phishing attacks and lowered the threshold for attackers. Modern phishing attacks now blend multiple tactics, including social engineering, URL spoofing, and AI deepfakes enabling adversaries to craft highly convincing messages that exploit human vulnerabilities and bypass traditional detection systems. At the same time, current security awareness education struggles to keep up with the speed, sophistication, and complexity of these evolving threats. To address this challenge, we propose a two-stage anti-phishing framework, CyberGLA, that combines technical defense and user-centered security education. In the Detection stage, we introduce EmailKnight, a spoof detection tool that performs multi-level email analysis. To enhance user awareness, the Training stage incorporates a large language model (LLM)-based security coach that dynamically selects personalized training modules based on the outcomes of the Detection stage. This dual purpose design philosophy enables effective protection against the evolving threats of modern email phishing attacks.

## I. Introduction

Email phishing remains one of the most persistent and damaging cyber threats. According to the report by Anti Phishing Working Group (APWG) [1], phishing attacks reached a new peak in the first quarter of 2024, with 1,003,924 incidents reported, the highest number since late 2023. Phishing is particularly dangerous as it often serves as the entry point for broader cyberattacks, including credential theft, ransomware, and large-scale fraud [2]. Despite decades of research and the deployment of spam filters [3], penetration tests [4], authentication protocols such as SPF, DKIM, and DMARC [5], and extensive user awareness campaigns [6], phishing attacks continue to cause significant financial and organizational losses [7].

Recent advances in artificial intelligence have further intensified this threat, generating two significant impacts on email phishing [8]: 1) it lowers the barrier for adversaries to initiate phishing attacks, and 2) it enables the creation of more sophisticated phishing emails that are increasingly difficult to detect. In 2024, it was reported that 4.7% of 386,000 analyzed malicious phishing emails were crafted using artificial intelligence [9]. With the assistance of AI tools, adversaries can now produce highly personalized, grammatically flawless phishing messages at scale. These messages seamlessly integrate traditional phishing techniques such as social engineering (exploiting human psychological vulnerabilities [10]), URL spoofing, open relay exploitation with AI-generated content [11]. Moreover, some AI agents are capable of simulating full conversations, mimicking writing styles or responding dynamically to users in phishing chat scenarios. These capabilities significantly enhance the credibility and effectiveness of phishing attempts, posing a growing threat to both individual users and organizations. Naturally, this emerging challenge gives rise to two critical questions:

1. What impact will AI-powered modern email phishing have on users, in particular, college students?
2. How can we effectively prevent and mitigate these advanced, AI-driven phishing attacks?

To address the two questions, in light of both the limitations of existing prevention approaches and the emerging challenges posed by AI-powered phishing explained

Published in JCISSE 13(1), Article 7 (2026). DOI: 10.53735/cisse.v13i1.230.

in Section II, we argue that combining technical defense (phishing email detection) with security education (user training) can significantly enhance the overall effectiveness of phishing prevention. Based on this insight, we propose CyberGLA, a two-stage anti-phishing framework that combines automated detection with personalized training. In the first Detection stage, EmailKnight analyzes incoming emails through multi-level inspection of headers, metadata, content, and attachments, extending beyond conventional rule-based protocols. In the second Training stage, CyberGLA delivers adaptive, scenario-specific lessons via an LLM-based security coach, translating detection results into targeted learning opportunities.

To evaluate the effectiveness of CyberGLA, we assessed the detection performance of EmailKnight, our phishing detection component, by comparing it with other widely available email phishing detection tools. This dual evaluation approach allowed us to measure both the real-world training impact of CyberGLA across diverse user demographics and the technical effectiveness of EmailKnight in identifying AI-powered phishing attacks. The details are discussed in Section IV. We also conducted a pilot study with 51 volunteer undergraduate participants, all of whom provided informed consent. Results from this small-scale evaluation suggest that students preferred the customized, scenario-driven content provided by CyberGLA and achieved higher overall quiz performance compared to baseline groups.

To summarize, our key contributions are as follows:

1. A robust phishing detection tool, EmailKnight, specifically designed to detect modern AI-enabled phishing threats.
2. A novel two-stage anti-phishing framework, CyberGLA, which integrates accurate email detection with personalized, LLM-driven security training.

## II. Background: Detection and Education

Phishing defenses generally fall into two domains: technical detection and user education.

*a) Detection:* Rule-based approaches [12] analyze meta-data and authentication protocols such as SPF, DKIM, and DMARC. Tools like SPFail evaluate SPF misconfigurations [13], while DKIM [14] and DMARC add validation but can fail under forwarding or misconfiguration. Our tool, EmailKnight, extends this category by incorporating human-factor cues, spoofed URLs, and AI-generated content. In parallel, machine learning and deep learning approaches classify emails using extracted features. Early models such as Random Forest [15] and Naïve Bayes [16] were followed by deep learning methods like CNNs [17], LSTMs [18], and NLP-based classifiers [19]. More recently, transformers [20] and LLM-based detectors [21] report up to 99% accuracy [22]. However, these models often struggle to generalize to real-world data as attackers increasingly exploit the same AI tools [23]. Hybrid systems are therefore needed for robust and adaptive detection. Incorporating deep learning based detection into CyberGLA's detection stage is one of the future work to improve the performance of our framework.

*b) Education:* Cybersecurity awareness training remains essential, but existing methods have limitations. Video-based training, offered by platforms such as Mimecast, Ninjio [24], remains widely used but often suffers from low engagement. Interactive training modules, such as embedded phishing simulations or scenario-based walkthroughs, have been shown to improve detection skills more effectively than static instruction [25], [26]. However, interactive simulations provide hands-on practice but can also lose impact over time. Gamified and interactive designs further enhance engagement and retention [27], [28]. However, studies also note that effectiveness varies with context and may decline without personalization [29]. Tools such as Spamley [30] provide quiz-based training similar to CyberGLA, though they lack coverage of advanced topics like open relay abuse and AI-generated content.

Together, these observations highlight the need for integrated solutions. Purely technical defenses may fail against evolving AI-powered attacks, while generic training struggles to sustain engagement. CyberGLA addresses this gap by combining detection (EmailKnight) with adaptive, scenario-specific education tailored to user vulnerabilities.

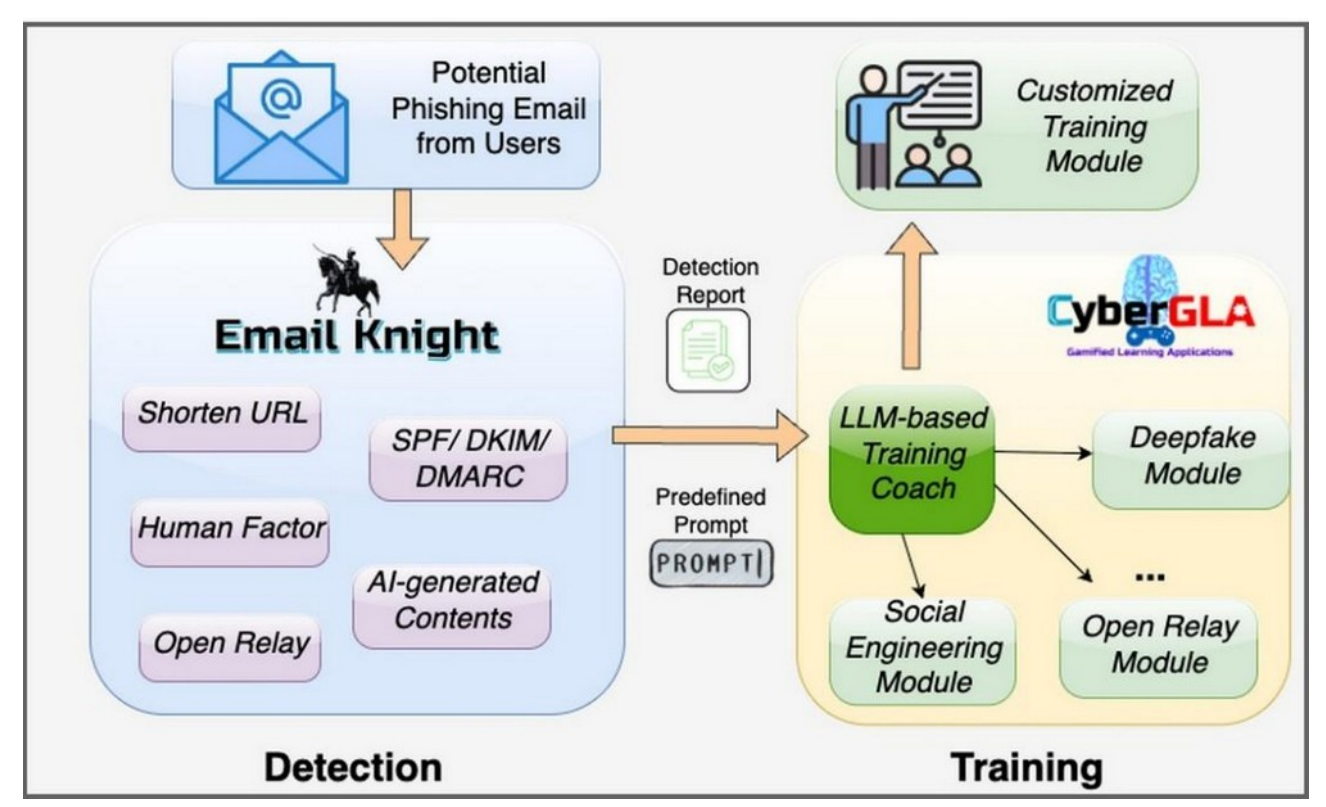

Fig. 1. The Overall Architecture of CyberGLA.

## III. System Architecture

The overall architecture of CyberGLA is shown in Figure 1. The detection component consists of EmailKnight, designed to detect modern email phishing powered by AI. EmailKnight analyzes suspicious aspects of potential phishing emails that the user has received, checking for issues such as SPF/DKIM/DMARC failures, the presence of shortened URLs, AI-generated content, persuasive or alarming language typical of traditional human factor phishing, or if the email originates from an open relay. The analysis results of the email will be made available to the user, allowing them to determine whether the email is trustworthy. Simultaneously, the analysis results are sent to our Training component, which includes pre-selected training modules for various modern phishing attacks and an LLM-based training coach. The training coach uses the analysis report, along with predefined prompts, to determine the most suitable training content to help the user understand the specific type of phishing email they received and reported to CyberGLA.

### A. Detection Tool: EmailKnight

EmailKnight is the detection component of CyberGLA, designed to identify phishing emails that evade conventional defenses. Unlike typical spam filters, it integrates protocol checks with behavioral and AI-based analysis, offering a multi-layered view of risk.

*1) Detection Capabilities:* EmailKnight integrates both technical and behavioral indicators. The following detection strategies are currently implemented:

1. **SPF/DKIM/DMARC Status Extraction:** The system parses `Authentication-Results` and `Received-SPF` headers to determine whether email authentication protocols have passed or failed. Failures, neutral outcomes, or missing fields trigger escalating threat scores.
2. **Open Relay Circumvention Detection:** Recognizes situations where emails pass all authentication checks but exhibit signs of spoofing, such as suspicious or forged device origin, anonymous headers, or domain mismatches, typically associated with open relay abuse.
3. **Header-Level Authentication Flags:** The tool inspects headers such as X-MS-Exchange-Organization-AuthAs, and X-MS-Exchange-CrossTenant-AuthAs. Messages labeled as "Anonymous" are flagged as elevated risk, while "Internal" values incur no penalty.
4. **Device Origin Analysis:** EmailKnight extracts `User-Agent`, `X-Mailer`, and `Received` headers to infer the sender's device environment. Suspicious indicators like "Desktop", "Laptop", or unrecognized clients are treated as red flags.
5. **Message-ID Anomaly Detection:** The domain inside the `Message-ID` is checked against the sender's claimed domain. Inconsistencies suggest potential spoofing or message injection.
6. **From/Reply-To/Return-Path Mismatches:** Mismatches between sender addresses and reply paths are evaluated. The system also flags discrepancies between displayed sender names and actual domains.
7. **URL Shortener and Link Display Checks:** Shortened URLs from known services (e.g., `bit.ly`, `tinyurl`) are flagged. Additionally, anchor tags in HTML content are analyzed for display mismatches between link text and the actual destination.
8. **Human Factors and Social Engineering Detection:** EmailKnight identifies manipulative language through a curated list of high-risk phrases (e.g., "urgent action required", "verify now"). To minimize false positives in academic and professional environments where individual urgency cues are common, the system utilizes a three-term threshold. While traditional Bayesian filters offer general spam classification, this targeted heuristic approach allows CyberGLA to specifically isolate behavioral human-factor cues often found in social engineering attacks without prematurely flagging legitimate administrative correspondence.
9. **AI-Driven Attachment Analysis:** Attachments including images, PDFs, videos, and audio files are scanned using pretrained ResNet and torchaudio models. Verdicts such as "Likely Real," "Possibly Cloned," or "Likely Fake" help evaluate if the content has been generated or manipulated using AI deepfake techniques.

*2) Usability and Privacy:* EmailKnight is designed for usability and privacy. All analysis runs locally on the user's device, without reliance on third-party APIs, ensuring compliance with institutional privacy requirements. The tool connects securely to mail servers via IMAP using app passwords and never stores sensitive content externally. From the user's perspective, suspicious emails can be scanned in real time and are assigned a color-coded threat score, accompanied by explanations of flagged anomalies as shown in Figure 2(a). Reports can be exported as PDF or CSV, supporting both individual users and institutional audits. A built-in quarantine option allows immediate isolation of risky messages, and user feedback (e.g., marking false positives) is incorporated into future scans to improve accuracy.

This architecture is particularly valuable for institutions that outsource cybersecurity or lack real-time forensic visibility. A growing concern is that many cyber insurance contracts or regulatory regimes allow **non-disclosure of minor breaches** if their impact falls below a financial or operational threshold (e.g., fewer than 500 individuals affected or no material harm) [31]. This creates a gap in visibility for individuals and organizations alike.

### B. Personalized Training Platform

CyberGLA's personalized training platform bridges the gap between technical threat detection and user awareness, turning each user's real inbox risks into scenario-based, adaptive security education. Rather than generic advice, CyberGLA creates a feedback loop: as EmailKnight scans emails and flags risky messages, those samples are used to shape the user's own learning modules.

*1) Integration of Detection and Training:* Whenever EmailKnight detects a potentially dangerous email, it is not only flagged but also categorized according to its underlying threat such as lookalike domains, open relay exploits, spoofed links, social engineering tactics, or even AI-generated deepfakes. For each flagged message, users receive an AI-generated summary that explains why the email is risky, what red flags are present, and specific next steps for remediation or reporting. Most importantly, CyberGLA tracks trends across a user's scans. If, for example, a user is repeatedly targeted by "urgent payment" scams or spoofed institutional messages, the platform automatically recommends a training module tailored to that threat type. This personalization ensures users focus on the attacks most relevant to them, instead of wading through irrelevant or generic cybersecurity content. In Figure 2(a), besides a high risk warning, CyberGLA recommends its interactive training module on human factors in phishing, presented in Figure 2(b), when the user faces a malicious email asking it to verify its login password of its bank account.

*2) Predefined Prompt and LLM Orchestration:* The "Predefined Prompt" in Figure 1 acts as the critical bridge between technical detection and pedagogical delivery. It serves as a system instruction for the LLM-based training coach, defining the rules for mapping technical anomalies from the Detection Report (e.g., SPF failures or open relay flags) to specific educational modules. By parsing these structured reports through the prompt, the coach dynamically generates a context-aware summary that explains the unique risks of the reported email, ensuring the subsequent training is directly relevant to the user's immediate experience.

*3) Modular, Scenario-Based Lessons:* Training is delivered through a growing library of modular, scenario-driven lessons, each focused on a major phishing technique. Modules include:

- **Human Factors in Phishing:** Covers social engineering, urgency, and trust cues commonly exploited in modern attacks.
- **Spotting Lookalike Domains:** Teaches users to identify typosquatting and visual deception in URLs and sender addresses.
- **Open Relay Attacks:** Explains how SMTP vulnerabilities can be used to bypass authentication protocols and impersonate trusted senders.
- **Spoofed Links:** Focuses on mismatched anchor text, shortened URLs, and link manipulation.
- **SPF/DKIM/DMARC Failures:** Describes authentication protocol bypasses and what warning signs to look for in email headers.
- **Recognizing Deepfakes:** Includes real video and audio deepfake demos to train users to spot AI-generated impersonation.

Each lesson is delivered through short, interactive scenarios such as quizzes, annotated email walkthroughs, or multimedia demonstrations. Unlike generic awareness programs, modules are personalized: repeated exposure to specific phishing types (e.g., urgent payment requests) triggers additional practice in those areas. Figure 2(b) shows one of the interactive learning modules on human factor topics. The platform is built as a modular web application, allowing rapid deployment of new scenarios as phishing evolves. The framework is designed with a flexible deployment architecture. While the current prototype is implemented as a modular web application, the system's decoupled backend is specifically architected for future integration as a browser plugin. Data handling is privacy-conscious: scans are processed securely, but no personal information is stored. By linking detection results to interactive, personalized lessons, CyberGLA transforms passive alerts into active learning. This approach addresses the limitations of static training and prepares users to recognize emerging phishing techniques, including those powered by AI.

*4) Interactive, Multimedia User Experience:* CyberGLA is designed for broad accessibility: there is no login or account creation required, and users can connect their inbox, upload an email file, or paste raw headers directly into the site for analysis. All results are delivered instantly and privately. After each scan, users see a simple dashboard with flagged emails, detailed threat explanations, and clear calls to action. The training modules themselves are interactive and engaging, and deepfake modules embed playable videos or audio clips. Scenario-based quizzes, and instant feedback help users test their recognition skills in a safe environment. This design supports experiential, hands-on learning rather than passive reading.

The platform's flexible architecture allows for future integration of features such as gamified progress tracking, badge systems or certificates of completion, and shareable learning reports as user needs evolve.

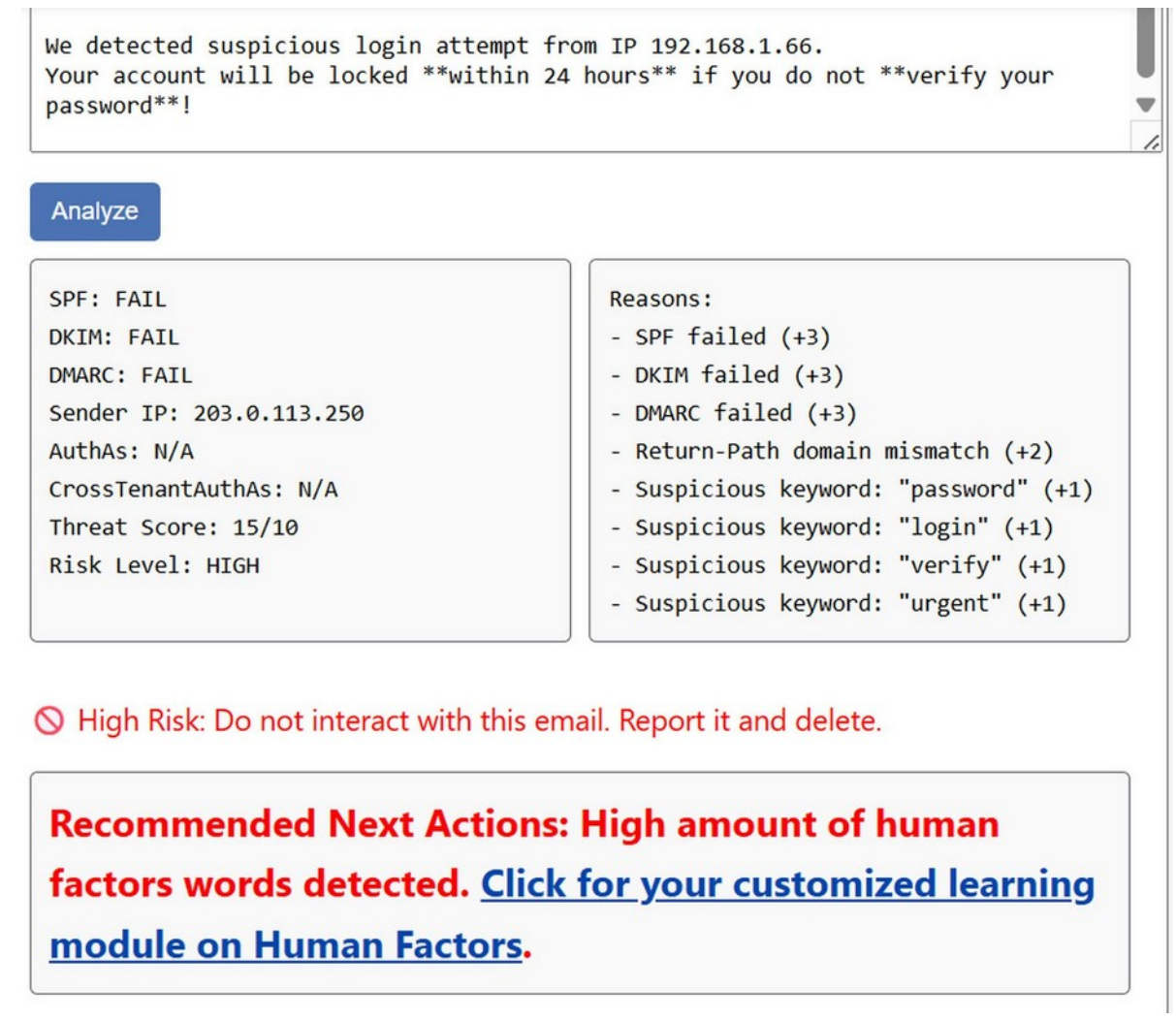

(a) EmailKnight's Result and CyberGLA's Customized Training Module Recommendation.

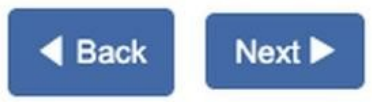

(b) Interactive Training Module on Human Factors.

Fig. 2. The Detection and Training Contents on CyberGLA's Web Application.

## IV. EVALUATION

### A. EmailKnight Evaluation

To assess the detection capabilities of EmailKnight, we conducted an empirical evaluation using five distinct phishing scenarios targeting Gmail, Yahoo, and university accounts. To ensure statistical significance and reflect modern attack volumes, we utilized an AI-assisted automation script to generate and transmit 100 unique phishing emails for each category (500 total malicious samples), along with a control group of benign emails. The five categories included: (1) open relay attacks with AI-generated templates and URL spoofing; (2) compromised account simulations featuring human-factor cues; (3) AI-generated attachments; (4) traditional human-factor phishing; and (5) benign correspondence. For each category, EmailKnight performed multi-level analysis, and the results presented in Table I represent the mean threat scores and classification accuracy across these 100-sample batches.

TABLE I. Evaluation Result of EmailKnight (Mean scores across $N$ = 500 automated test cases)

| Tool | Email Knight | MailWasher | MX ToolBox |
|---|---|---|---|
| Test 1 | High risk(7) | No flag | DKIM |
| Test 2 | Moderate risk(4) | Showed Real URL | No flag |
| Test 3 | Moderate risk(5) | No flag | DMARC |
| Test 4 | Moderate risk(4) | No flag | DKIM |
| Test 5 | Low risk(2) | No flag | No flag |

We compared EmailKnight against MailWasher [32] and MXToolbox [33]. Results are shown in Table I. EmailKnight flagged all phishing cases and correctly classified the benign email. MXToolbox provided header details (e.g., DKIM, DMARC) but no actionable recommendations and failed to detect human-factor cues, open relays, or AI content. MailWasher revealed spoofed URLs but did not flag phishing or suggest remediation. We used its free version. We also tested Cleanfox, which required app passwords and permissions but offered only inbox organization rather than phishing detection, so it was excluded from comparison.

### B. Education Effective Evaluation

*1) Setting-up:* We evaluated CyberGLA through a controlled pilot study with 51 participants, including undergraduate and graduate students at a mid-sized U.S. university. Participants were recruited through class announcements, and faculty outreach. All participants are provided informed electronic consent in the case study. Data collection included online anonymous quiz performance, and optional anonymous feedback, and no personal identifiers were stored. The study was conducted under the university's privacy and ethics policies.

Participants were randomly divided into three groups. One trained with CyberGLA(18 members), while the other trained with Mimecast (18 members), a widely used commercial awareness platform, the baseline group with no training (15 members). After 30 minutes of learning, all groups took a detailed online quiz of 25 multiple choice questions on various aspects of email phishing 25 multiple-choice questions designed to assess the effectiveness of the training methods.

*2) Results:*

*a) Quantitative Results:* Participants trained with CyberGLA outperformed those trained with Mimecast. Average quiz accuracy improved by 20–25% in the CyberGLA group, compared to modest gains in the Mimecast group as shown in Table II. Many feedback notes showed that interactive, scenario-driven modules were more memorable than static video lessons. This finding is consistent with prior research showing that game-based and interactive training can improve phishing awareness and user engagement [34], [35].

TABLE II. Evaluation Result of Groups with Training.

| Group | CyberGLA | Mimecast | No training |
|---|---|---|---|
| Avg score | 93.33 | 85.33 | 82 |

*b) Qualitative Results:* Out of the 18 participants in the CyberGLA group, 14 (78%) submitted anonymous on-line feedback. All respondents indicated that CyberGLA was helpful and that they would recommend it (via two selection buttons on the feedback form). Representative positive comments included "Very fun" and "I think it will be helpful for kids and elderly people." Participants also offered constructive suggestions, such as "I would like it to cover more details" and "I would like to play games in these learning modules." This indicates the potential of improving the engagement of customized training in security education, aligning with the design goal of CyberGLA.

Overall, this pilot study suggests that combining detection with adaptive, customized training can improve phishing awareness and engagement compared to a commercial base-line. However, these findings should be viewed as exploratory evidence rather than definitive proof. A larger-scale, more diverse, and longitudinal evaluation will be necessary to validate CyberGLA's effectiveness beyond the academic context.

*3) Limitation and Threats:* This pilot study has several limitations. The sample size was modest, restricting statistical generalizability. First, while the EmailKnight component was tested against a substantial sample of 500 malicious emails ($N$ = 100 per category) using AI-assisted automation scripts, these samples were generated through specific templates. Consequently, the results may not fully capture the evolving obfuscation techniques used by diverse adversarial groups, and the modest variety of scenarios restricts broader technical generalizability. This challenge is not unique: designing rigorous phishing studies is inherently difficult due to ecological validity and participant bias [36]. Second, the participant pool was confined to an academic setting, where phishing scenarios often reference coursework or institutional branding. Results may not extend to corporate or high-stakes environments. Additionally, the study was short-term, and long-term knowledge retention was not assessed.

### *C. Self-Critique and Future Considerations*

CyberGLA faces challenges in scaling to real world deployments. Its cultural and contextual adaptability remains untested, as phishing cues vary across regions and organizations. Also, EmailKnight's detection modules, which incorporate curated phishing phrase lists, must be updated frequently to reflect evolving attacker language. Automating this process using phishing corpus mining or integration with threat intelligence feeds is a future priority.

## V. Ethics and Human Subjects Research

This study involved human participants in phishing simulations and personalized security training. All procedures were conducted in accordance with the university's privacy policy and human subjects research guidelines.

*a) Informed Consent:* The 51 students volunteered for the evaluation with oral and online electronic consent. Participation was strictly voluntary, and no incentives were provided.

*b) Data Collection and Anonymity:* No personally identifying information was collected. In the evaluation, only aggregate, non-identifiable data (scores, and optional anonymous feedback) were logged. No institutional login data, IP addresses, or demographic identifiers were retained.

c*) Recruitment and Oversight:* Participants were recruited through classroom announcements and departmental outreach. No student's grade, standing, or enrollment was affected. The Department consulted the Institutional Review Board (IRB), which confirmed the study fell within low-risk educational research and did not require full review.

*d) Risk Mitigation:* All simulated phishing experiments were debriefed with participants, clarifying that no real accounts had been compromised. All experiments were designed to maintain minimal risk and educational benefit for participants.

## VI. Conclusions and Future Work

This paper presented a two-stage anti-phishing framework that integrates technical detection with adaptive, customized interactive training. Future work will focus on expanding the empirical evaluation to larger and more diverse populations, specifically utilizing standardized phishing corpuses to provide higher statistical significance regarding detection efficacy. We also aim to explore longitudinal impacts and integrate automated updates to the phrase bank using phishing corpus mining and threat intelligence feeds. Beyond email phishing, we intend to address other emerging electronic crimes, including advanced text phishing, voice cloning phone scams, and video call deepfakes. A primary goal is to adapt these scenarios for high-risk demographics, particularly the elderly population, who are increasingly targeted by such sophisticated techniques. Finally, based on participant feedback regarding the motivational benefits of gamification, we plan to further develop interactive games for additional training modules to enhance user engagement.

## VII. Acknowledge

This work is supported by the National Science Foundation (NSF) under Grant No. CRII 2451348.

## References

[1] apwg, "Phishing activity trends reports (1st quarter 2025)," 2025. [Online]. Available: https://apwg.org/trendsreports/

[2] E. H. Tusher, M. A. Ismail, M. A. Rahman, A. H. Alenezi, and M. Uddin, "Email spam: A comprehensive review of optimize detection methods, challenges, and open research problems," *IEEE Access*, vol. 12, p. 143627–143657, 2024. [Online]. Available: http://dx.doi.org/10.1109/access.2024.3467996

[3] N. J. Palatty, "81 phishing attack statistics 2025: The ultimate insight," 2025. [Online]. Available: https://www.getastra.com/blog/security-audit/phishing-attack-statistics/

[4] E. Gardner, G. Singh, and W. Qu, "Penetration testing operating systems: Exploiting vulnerabilities," in *2024 International Conference on Communications, Computing, Cybersecurity, and Informatics (CCCI)*. IEEE, Oct. 2024, p. 1–9. [Online]. Available: http://dx.doi.org/10.1109/ccci61916.2024.10736454

[5] S. C. Sethuraman, D. P. V S, T. Reddi, M. S. T. Reddy, and M. K. Khan, "A comprehensive examination of email spoofing: Issues and prospects for email security," *Computers & Security*, vol. 137, p. 103600, Feb. 2024. [Online]. Available: http://dx.doi.org/10.1016/j.cose.2023.103600

[6] S. Cook, "Phishing email attack statistics and facts for 2019–2024," 2024. [Online]. Available: https://www.comparitech.com/blog/vpn-privacy/phishing-statistics-facts/

[7] U. A. Office, "Indictment charges two in $230 million cryptocurrency scam," 2025. [Online]. Available:

https://www.justice.gov/usao-dc/pr/indictment-charges-two-230-million-cryptocurrency-scam

[8] N. Marshall, D. Sturman, and J. C. Auton, “Exploring the evidence for email phishing training: A scoping review,” *Computers & Security*, vol. 139, p. 103695, Apr. 2024. [Online]. Available: http://dx.doi.org/10.1016/j.cose.2023.103695

[9] M. Cartier, “Ai phishing attacks: How big is the threat?” 2025. [Online]. Available: https://hoxhunt.com/blog/ai-phishing-attacks

[10] L. Gallo, D. Gentile, S. Ruggiero, A. Botta, and G. Ventre, “The human factor in phishing: Collecting and analyzing user behavior when reading emails,” *Computers & Security*, vol. 139, p. 103671, Apr. 2024. [Online]. Available: http://dx.doi.org/10.1016/j.cose.2023.103671

[11] K. A. Pantserev, *The Malicious Use of AI-Based Deepfake Technology as the New Threat to Psychological Security and Political Stability*. Springer International Publishing, 2020, p. 37–55. [Online]. Available: http://dx.doi.org/10.1007/978-3-030-35746-7_3

[12] J. Berkowitz and W. Qu, *A Static Over-Approximate Detection Tool for At-Risk DLLs*. Springer Nature Switzerland, 2025, p. 516–525. [Online]. Available: http://dx.doi.org/10.1007/978-3-031-86637-1_38

[13] N. Bennett, R. Sowards, and C. Deccio, “Spfail: discovering, measuring, and remediating vulnerabilities in email sender validation,” in *Proceedings of the 22nd ACM Internet Measurement Conference*, ser. IMC ’22. ACM, Oct. 2022, p. 633–646. [Online]. Available: http://dx.doi.org/10.1145/3517745.3561468

[14] M. Thomas, *Requirements for a DomainKeys Identified Mail (DKIM) Signing Practices Protocol*, Oct. 2007. [Online]. Available: http://dx.doi.org/10.17487/rfc5016

[15] A. A. Akinyelu and A. O. Adewumi, “Classification of phishing email using random forest machine learning technique,” *Journal of Applied Mathematics*, vol. 2014, p. 1–6, 2014. [Online]. Available: http://dx.doi.org/10.1155/2014/425731

[16] K. Agarwal and T. Kumar, “Email spam detection using integrated approach of naïve bayes and particle swarm optimization,” in *2018 Second International Conference on Intelligent Computing and Control Systems (ICICCS)*. IEEE, Jun. 2018, p. 685–690. [Online]. Available: http://dx.doi.org/10.1109/iccons.2018.8662957

[17] R. Alotaibi, I. Al-Turaiki, and F. Alakeel, “Mitigating email phishing attacks using convolutional neural networks,” in *2020 3rd International Conference on Computer Applications & Information Security (ICCAIS)*. IEEE, Mar. 2020, p. 1–6. [Online]. Available: http://dx.doi.org/10.1109/iccais48893.2020.9096821

[18] Q. Li, M. Cheng, J. Wang, and B. Sun, “Lstm based phishing detection for big email data,” *IEEE Transactions on Big Data*, vol. 8, no. 1, p. 278–288, Feb. 2022. [Online]. Available: http://dx.doi.org/10.1109/tbdata.2020.2978915

[19] S. Salloum, T. Gaber, S. Vadera, and K. Shaalan, “Phishing email detection using natural language processing techniques: A literature survey,” *Procedia Computer Science*, vol. 189, p. 19–28, 2021. [Online]. Available: http://dx.doi.org/10.1016/j.procs.2021.05.077

[20] R. Mel éndez, M. Ptaszynski, and M. Fumito, “Comparative investigation of traditional machine learning models and transformer models for phishing email detection,” Nov. 2024. [Online]. Available: http://dx.doi.org/10.20944/preprints202410.1467.v2

[21] M. A. Uddin and I. H. Sarker, “An explainable transformer-based model for phishing email detection: A large language model approach,” 2024. [Online]. Available: http://dx.doi.org/10.2139/ssrn.4785953

[22] A. Alhuzali, A. Alloqmani, M. Aljabri, and F. Alharbi, “In-depth analysis of phishing email detection: Evaluating the performance of machine learning and deep learning models across multiple datasets,” *Applied Sciences*, vol. 15, no. 6, p. 3396, Mar. 2025. [Online]. Available: http://dx.doi.org/10.3390/app15063396

[23] C. S. Eze and L. Shamir, “Analysis and prevention of ai-based phishing email attacks,” *Electronics*, vol. 13, no. 10, p. 1839, May 2024. [Online]. Available: http://dx.doi.org/10.3390/electronics13101839

[24] Ninjio, “Reduce human based cyber risk,” 2025. [Online]. Available: https://ninjio.com/

[25] P. Kumaraguru, S. Sheng, A. Acquisti, L. F. Cranor, and J. Hong, “Teaching johnny not to fall for phish,” *ACM Transactions on Internet Technology*, vol. 10, no. 2, p. 1–31, May 2010. [Online]. Available: http://dx.doi.org/10.1145/1754393.1754396

[26] M. Alsharnouby, F. Alaca, and S. Chiasson, “Why phishing still works: User strategies for combating phishing attacks,” *International Journal of Human-Computer Studies*, vol. 82, p. 69–82, Oct. 2015. [Online]. Available: http://dx.doi.org/10.1016/j.ijhcs.2015.05.005

[27] J.-N. Tioh, M. Mina, and D. W. Jacobson, “Cyber security training a survey of serious games in cyber security,” in 2017 *IEEE Frontiers in Education Conference (FIE)*. IEEE, Oct. 2017, p. 1–5. [Online]. Available: http://dx.doi.org/10.1109/fie.2017.8190712

[28] P. Weanquoi, J. Johnson, and J. Zhang, “Using a game to improve phishing awareness,” *Journal of Cybersecurity Education, Research and Practice*, vol. 2018, no. 2, Dec. 2018. [Online]. Available: http://dx.doi.org/10.62915/2472-2707.1040

[29] C. I. Canfield, B. Fischhoff, and A. Davis, “Quantifying phishing susceptibility for detection and behavior decisions,” *Human Factors: The Journal of the Human Factors and Ergonomics Society*, vol. 58, no. 8, p. 1158–1172, Sep. 2016. [Online]. Available: http://dx.doi.org/10.1177/0018720816665025

[30] F. Pietrantonio, A. Botta, S. Zinno, G. Ventre, L. Gallo, L. Mancuso, and R. Presta, “A gaze-based analysis of human detection of email phishing,” in *2024 Silicon Valley Cybersecurity Conference (SVCC)*. IEEE, Jun. 2024, p. 1–8. [Online]. Available: http://dx.doi.org/10.1109/svcc61185.2024.10637355

[31] C. I. S. Agency, “Cyber incident reporting for critical infrastructure act (circia),” 2022. [Online]. Available: https://www.cisa.gov/circia

[32] MailWasher, “Mailwasher,” 2025. [Online]. Available: https://mailwasher.net/

[33] MXtoolbox, “Mxtoolbox supertool,” 2025. [Online]. Available: https://mxtoolbox.com/SuperTool.aspx

[34] S. Sheng, B. Magnien, P. Kumaraguru, A. Acquisti, L. F. Cranor, J. Hong, and E. Nunge, “Anti-phishing phil: the design and evaluation of a game that teaches people not to fall for phish,” in *Proceedings of the 3rd symposium on Usable privacy and security*, ser. SOUPS ’07. ACM, Jul. 2007, p. 88–99. [Online]. Available: http://dx.doi.org/10.1145/1280680.1280692

[35] N. A. G. Arachchilage and S. Love, “A game design framework for avoiding phishing attacks,” *Computers in Human Behavior*, vol. 29, no. 3, p. 706–714, May 2013. [Online]. Available: http://dx.doi.org/10.1016/j.chb.2012.12.018

[36] K. Parsons, A. McCormac, M. Pattinson, M. Butavicius, and Jerram, “The design of phishing studies: Challenges for researchers,” *Computers & Security*, vol. 52, p. 194–206, Jul. 2015. [Online]. Available: http://dx.doi.org/10.1016/j.cose.2015.02.008